\documentclass[twocolumn,prl,nofootinbib,preprintnumbers,superscriptaddress]{revtex4}

\usepackage{graphicx}
\usepackage{multirow}
\usepackage{enumitem}
\usepackage{amssymb}
\usepackage{amsmath}
\usepackage{xcolor}
\usepackage{xspace}
\usepackage[normalem]{ulem} 

\usepackage{hyperref}

\usepackage{booktabs} 

\newcommand{\beq}{\begin{equation}}
\newcommand{\eeq}{\end{equation}}
\newcommand{\beqa}{\begin{eqnarray}}
\newcommand{\eeqa}{\end{eqnarray}}

\def\OMIT#1{{}}

\makeatletter
\g@addto@macro\bfseries{\boldmath}
\let\Hy@backout\@gobble
\makeatother

\begin{document}
\preprint{CERN-TH-2025-077}

\title{Natural complex plane for kaon CKM data: framework, status and future}

\author{Avital Dery}
\email{avital.dery@cern.ch}
\affiliation{CERN, Theoretical Physics Department, Geneva, Switzerland}

\begin{abstract}
    Kaon physics can be used to independently determine three out of the four parameters of the CKM matrix, without any B physics input. Treating one parameter, $|V_{us}|$, or alternatively Wolfenstein $\lambda$, as well known, we show that the natural plane for the presentation of kaon CKM information is spanned by the combinations $\left(A^2(1-\hat\rho),\, A^2 \hat\eta\right)$. 
    In this way, the use of B physics inputs is avoided, as well as the artificial inflation of errors due to parametric uncertainties, mainly due to $|V_{cb}|$.
    We show that the current status of kaon CKM constraints, impacted by recent advances in measurement and theory, is characterized by four allowed regions, and find that incoming data will inevitably disfavor a number of them, either confirming the CKM paradigm as dominant, or discovering a departure from the Standard Model.
\end{abstract}

\maketitle

{\bf Introduction}\quad
Recent advances in both experiment and theory of rare kaon decays~\cite{NA62:2024pjp,Hoferichter:2023wiy,DAmbrosio:2017klp,Dery:2021mct,Dery:2022yqc,Chao:2024vvl,KOTO:2025gvq,Dery:ToAppear} allow to imagine the use of pure kaon physics information to independently determine parameters of the Cabibbo-Kobayashi-Maskawa (CKM) matrix.
The CKM matrix governs flavor changing processes as well as CP violation (CPV) within the Standard Model (SM) of particle physics. The determination of its parameters has so far been dominated by constraints involving B mesons. 
Current and expected knowledge from the kaon sector, providing internal consistency checks as well as testing coherence with the B physics picture, would amount to a unique and crucial test of the CKM paradigm;
This would be a cross check of the CKM mechanism across sectors, probing the viable possibility that physics beyond the SM (BSM) affects kaons and B mesons differently.

The CKM matrix is a $3\times 3$ unitary matrix, 
\begin{equation}
    \{V_{ij}\}\, \qquad  i=u,c,t\,, \qquad j=d,s,b\, ,
\end{equation}
connecting between the mass bases of the up- and down- type quarks. 
The orthogonality of any pair of its rows or columns implies six closure conditions of the form
\begin{equation}
    \sum_{i=u,c,t} V_{iq}V_{iq^\prime}^*=0 \, ; \qquad \sum_{j=d,s,b} V_{qj}V_{q^\prime j}^*=0\, ,
\end{equation}
which are referred to as \textit{unitarity triangles}.
Taking into account the unitarity constraints as well as the symmetries of the SM lagrangian, the CKM is characterized by four real physical parameters, that correspond to three magnitudes and one phase. In the Wolfenstein convention~\cite{PhysRevLett.51.1945}, these are denoted by $\{A,\,\lambda,\,\rho,\,\eta\}$, where $\lambda$ is a small parameter and $\eta$ controls the imaginary parts.

The cross determination of CKM parameters from various measurements is a test of the SM, so far holding beautifully under scrutiny of a collection of B physics precision observables.
An orthogonal determination from the kaon sector, completely independent of B physics inputs, has been one of the main ambitions of the kaon program~\cite{Buchalla:1996fp,Buras:2006gb,Blucher:2009zz,Lehner:2015jga,Buras:2021nns,Lunghi:2024sjy}, and is now becoming within reach.

The analysis and presentation of information from the kaon sector requires, however, methods that are specifically suited to it, rather than borrowed from the existing B sector framework.
The dominance of B physics constraints in the past decades has dictated the traditional presentation of consistency between observables, in the complex plane given by~\cite{Buras:1994ec,Charles:2004jd},
\begin{equation}
    \bar\rho + i\bar\eta \, \equiv \, R_u \,e^{i\gamma}\, ,
\end{equation}
where $R_u$ and $\gamma$ are a side and an angle of the normalized $(bd)$ unitarity triangle,
\begin{equation}
    R_u = \left|\frac{V_{ud}V_{ub}^*}{V_{cd} V_{cb}^*}\right|\, ,\qquad \gamma=\arg\left(-\frac{V_{ud}V_{ub}^*}{V_{cd}V_{cb}^*}\right)\, ,
\end{equation}
and $(\bar\rho,\bar\eta)$ coincide with $(\rho,\eta)$ to ${\cal O}(\lambda^2)$.
\begin{figure}[t!]
    \centering
    \includegraphics[width=1.0\linewidth]{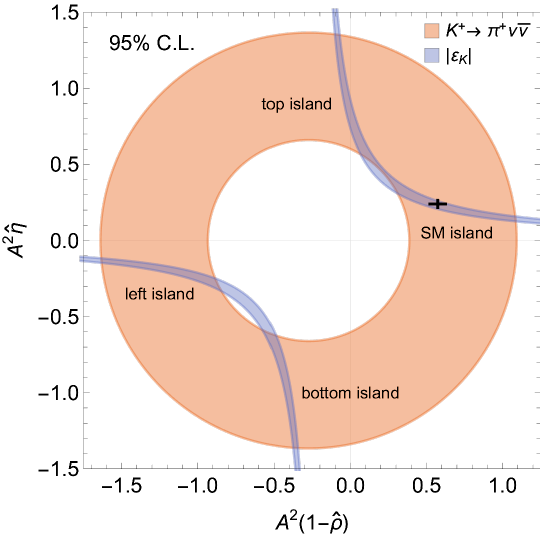}
    \caption{Current allowed regions from measurements of ${\cal B}(K^+\to\pi^+\nu\bar\nu)$ and $|\varepsilon_K|$ at $95\%$ confidence level. The SM global fit point~\cite{ParticleDataGroup:2024cfk} (B physics dominated) is marked in black. Four almost disconnected regions are allowed by the conjunction of these two measurements, denoted here as the ``left", ``top", ``bottom" and ``SM" islands.}
    \label{fig:2sigmaCurrent}
\end{figure}

However, for kaon observables, the use of the $\left(\bar\rho,\,\bar\eta\right)$ plane makes a pure kaon physics determination (free of B physics inputs) impossible, 
since kaon constraints are dependent on a different parameter combination, that involves factors of $A$.
Projected on the $\left(\bar\rho,\,\bar\eta\right)$ plane, this then
implicitly implies that the parameter $A$, or $|V_{cb}|$, must be used as input. 
We note that the parameter $\lambda$, related to the magnitude $|V_{us}|$, is determined from kaon physics, and for the purposes of this analysis we treat it as well known. 
On the other hand, the parameter $A$, derived from the magnitude $|V_{cb}|\approx A\lambda^2$, is determined from semileptonic B decays, and cannot be determined by kaon physics (excluding a possible future determination of $|V_{ts}|$ from top decays involving kaons).
The use of $|V_{cb}|$ as input in kaon constraints therefore not only inflates the apparent errors due to the currently sizable parametric uncertainty, but more importantly stands in the way of an independent kaon physics picture 
(for other approaches to eliminate $|V_{cb}|$ dependence by marginalization, see, for example, Refs.~\cite{Buras:2021nns,Lunghi:2024sjy}).

Here we argue that, for the purposes of kaon CKM information, the natural choice of the complex plane of interest is rather given by
\begin{equation}\label{eq:rhohatetahatDef}
     A^2 \lambda^4  (1-\hat\rho+i\hat\eta) \, \equiv \, -R_{ct} \,e^{-i\theta_{ct}}\, ,
\end{equation}
where  $R_{ct}$ and $\theta_{ct}$ are a side and an angle of the normalized $(ds)$ unitarity triangle~\cite{Lebed:1996ay},
\begin{equation}
    R_{ct} = \left|\frac{V_{td}V_{ts}^*}{V_{cd}V_{cs}^*}\right|\, , \quad \theta_{ct} = \arg\left(-\frac{V_{td}V_{ts}^*}{V_{cd}V_{cs}^*}\right)\, .
\end{equation}
We note that Eq.~\eqref{eq:rhohatetahatDef} is the definition of the parameters $(\hat\rho,\hat\eta)$, which coincide to ${\cal O}(\lambda^2)$ with $(\bar\rho,\bar\eta)$.

In comparison with the $\left(\bar\rho,\,\bar\eta\right)$ plane of B physics, the $\left(A^2\lambda^4(1-\hat\rho),\,\, A^2\lambda^4\hat\eta\right)$ plane differs by
\begin{enumerate}
    \item An overall scaling by a factor of $A^2\lambda^4$, common to all $(V_{td}V_{ts}^*)/(V_{cd}V_{cs}^*)$-dependent kaon observables, 

    \item A transformation of the $x$-axis, $\bar\rho \to (1-\hat\rho)$.
\end{enumerate}
In the following, and in all plots, we omit factors of $\lambda$ and treat it as known, substituting the PDG global fit value~\cite{ParticleDataGroup:2024cfk}. 

\begin{figure*}[ht]
    \centering
    \includegraphics[width=1.0\linewidth]{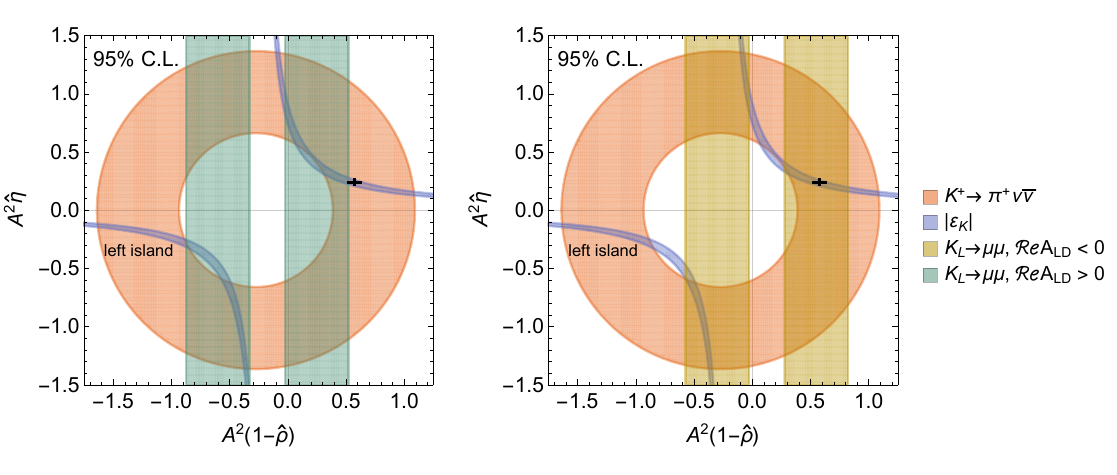}
    \caption{The addition of current ${\cal B}(K_L\to\mu^+\mu^-)$ information to the allowed regions of Fig.~\ref{fig:2sigmaCurrent}. The prediction of ${\cal B}(K_L\to\mu^+\mu^-)$ includes a fourfold discrete ambiguity, depending on the sign of the real part of the long-distance amplitude, ${\cal R}eA_{\rm LD}$, and on the relative sign of the short-distance amplitude. The left and right plots correspond to the interpretation of the measurements with positive and negative ${\cal R}eA_{\rm LD}$, respectively. The ``left island" is marginally ruled out at $95\%\,\,{\rm C.L.}$ if ${\cal R}eA_{\rm LD}>0$ (left) and ruled out completely if ${\cal R}eA_{\rm LD}<0$ (right).}
    \label{fig:2sigmaKLmumu}
\end{figure*}
%

\begin{figure*}[ht]
    \centering
    \includegraphics[width=1.0\linewidth]{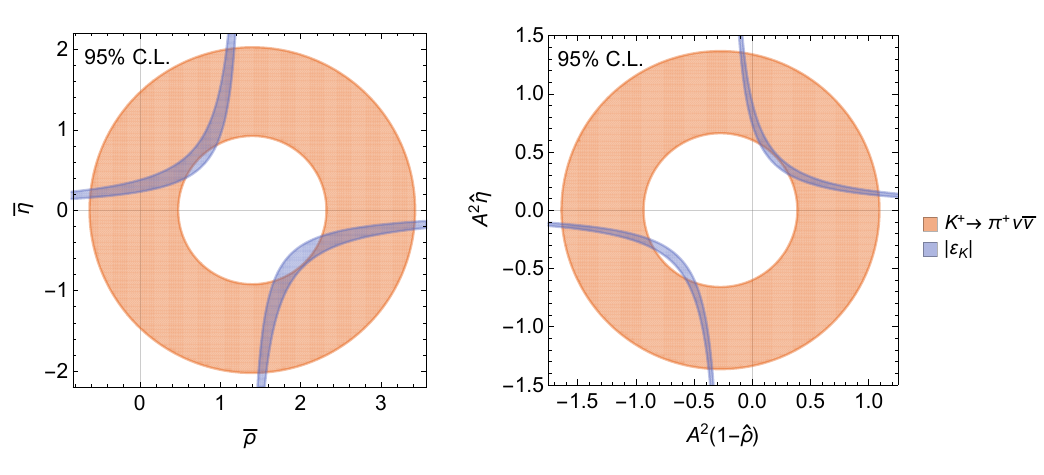}
    \caption{Supplemental Figure: The two kaon constraints, ${\cal B}(K^+\to\pi^+\nu\bar\nu)$ and $|\varepsilon_K|$, as depicted in Fig.~\ref{fig:2sigmaCurrent} (right) vs. the same constraints presented on the $(\bar\rho,\bar\eta)$ plane (left). The $|\varepsilon_K|$ bands show significant broadening on the left, due to artificial inflation of errors from the parametric uncertainty on the parameter $A$ (the same occurs for ${\cal B}(K^+\to\pi^+\nu\bar\nu)$, in principle, however the uncertainty is currently experimentally dominated). We emphasize that for the left plot B-physics input cannot be avoided, while the right plot is independent of B constraints. }
    \label{fig:RhoEta}
\end{figure*}

\vskip0.5cm
{\bf Current Kaon Parameter Space}\quad
To demonstrate the advantage of the use of the $\left(A^2(1-\hat\rho),\,\, A^2\hat\eta\right)$ plane, we first present the overlap of two leading kaon CKM measurements: the rate of the rare decay $K^+\to\pi^+\nu\bar\nu$, recently measured  by the NA62 collaboration~\cite{NA62:2024pjp}, and the kaon mixing parameter $|\varepsilon_K|$~\cite{ParticleDataGroup:2024cfk}.
Figure~\ref{fig:2sigmaCurrent} presents the current knowledge from these two observables, together with the SM point as taken from the PDG global fit~\cite{ParticleDataGroup:2024cfk}.
The strength of this approach is twofold --- first, no B physics CKM inputs are used, and second, the allowed regions are no longer inflated by the uncertainty on the parameter $A$, as is the case, for comparison, in the CKMFitter plot~\cite{CKMFitter}, where the $|V_{cb}|^4$ dependence of $|\varepsilon_K|$ implies wider bands. Instead, here $|\varepsilon_K|$ is viewed as a clean (sub $5\%$) measurement of this specific parameter combination (see Figure~\ref{fig:RhoEta} for a comparison of the same kaon constraints on the two planes).

The intersection of the two constraints features four (almost) disconnected allowed regions, which we denote as the: ``SM-", ``left-", ``top-", and ``bottom-" islands.
We note that the circular ring of the NA62 measurement is centered at $\hat\eta=0$, but in relation to the $x$-axis it is shifted from $(1-\hat\rho)=0$ due to the charm-loop contribution (see the Appendix for detailed expressions). Incidentally, a similar but not identical shift exists for the hyperbolic bands derived by the measurement of $|\varepsilon_K|$. This results in the approximate symmetry exhibited by the four islands in Fig.~\ref{fig:2sigmaCurrent}.

\vskip0.5cm
{\bf $K_L\to\mu^+\mu^-$: $x$-Axis Information}\quad
Additional relevant information can be obtained by examining the decay $K_L\to\mu^+\mu^-$, using the recently improved SM prediction~\cite{Hoferichter:2023wiy}.
The measurement of ${\cal B}(K_L\to\mu^+\mu^-)$ can be thought of as a measurement of 
\begin{equation}
    -R_{ct}\cos\theta_{ct} = A^2\lambda^4(1-\hat\rho)\, 
\end{equation}
(see detailed expressions in the Appendix).
Figure~\ref{fig:2sigmaKLmumu} shows the overlap of the two interpretations of the measurement of ${\cal B}(K_L\to\mu^+\mu^-)$ with the constraints shown in Fig.~\ref{fig:2sigmaCurrent}.
This mode is dominated by long-distance non-perturbative effects, which is reflected in the relatively large theory uncertainty, as well as in a sign ambiguity. Nevertheless, the allowed bands show significant tension with almost the entire ``left island".
This motivates further progress on the theory side --- if the uncertainty in the prediction of ${\cal B}(K_L\to\mu^+\mu^-)$ were to reduce, the "left island" could be excluded with ${\rm C.L.}>95\%$ in the near future. 
This provides additional motivation to improve the SM prediction, with prospects for advancements both using dispersion theory~\cite{Hoferichter:2023wiy} and Lattice QCD~\cite{Chao:2024vvl}.

\begin{figure}[ht]
    \centering
    \includegraphics[width=1.0\linewidth]{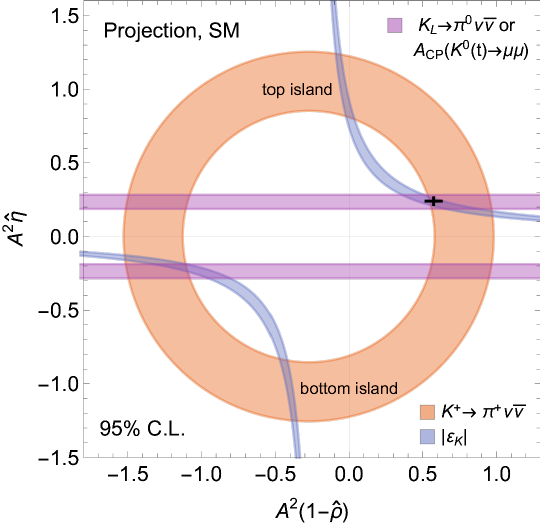}
    \caption{Future projection, assuming $15\%$ uncertainty on the measurement of ${\cal B}(K^+\to\pi^+\nu\bar\nu)$ (with current central value), and $20\%$ uncertainty on ${\cal B}(K_L\to\pi^0\nu\bar\nu)$ (or its equivalent), assuming agreement with the SM. }
    \label{fig:FutureSM}
\end{figure}
%
\vskip0.5cm
{\bf Upcoming $y$-Axis Information}\quad
Two complementary probes, measuring
\begin{equation}\label{eq:RctsinThetact}
    R_{ct}\sin\theta_{ct}
    = A^2\lambda^4\hat\eta\, , 
\end{equation}
are the future measurements of the rare decay ${\cal B}(K_L\to\pi^0\nu\bar\nu)$ and the CP asymmetry, $A_{\rm CP}(K^0\to\mu^+\mu^-)(t)$. The first is the main goal of the planned KOTO-II experiment~\cite{KOTO:2025gvq}, and the second is being investigated in the framework of an LHCb-like setup~\cite{Dery:ToAppear} (see also the study of $K(t)\to\mu^+\mu^-$ in the context of a proposal for a dedicated experiment~\cite{Marchevski:2023kab}). 
CPV in $K\to\mu^+\mu^-$ and ${\cal B}(K_L\to\pi^0\nu\bar\nu)$ would both be measurements of the parameter combination in Eq.~\eqref{eq:RctsinThetact} within the SM, and have complementary sensitivity to BSM operators~\cite{Dery:2021vql}.

Figure~\ref{fig:FutureSM} shows future projections, using estimations by the KOTO-II and NA62 collaborations ($15\%$ error on ${\cal B}(K^+\to\pi^+\nu\bar\nu)$ and $20\%$ for ${\cal B}(K_L\to\pi^0\nu\bar\nu)$), assuming the current central value for the former and agreement with the SM for the latter.
As can be seen, once SM sensitivity is reached for ${\cal B}(K_L\to\pi^0\nu\bar\nu)$ or $A_{\rm CP}(K^0\to\mu^+\mu^-)$ (even before the target of $20\%$ uncertainty is met), these measurements will either be in significant tension with the SM, or will strongly disfavor the ``top" and ``bottom" islands. 
In conjunction with the information from ${\cal B}(K_L\to\mu^+\mu^-)$, assuming reduced theory errors, this future situation would leave only the "SM island" as the allowed region from kaon physics, or alternatively discover a departure from the SM.

\begin{figure*}[ht!]
    \centering
    \includegraphics[width=0.9\linewidth]{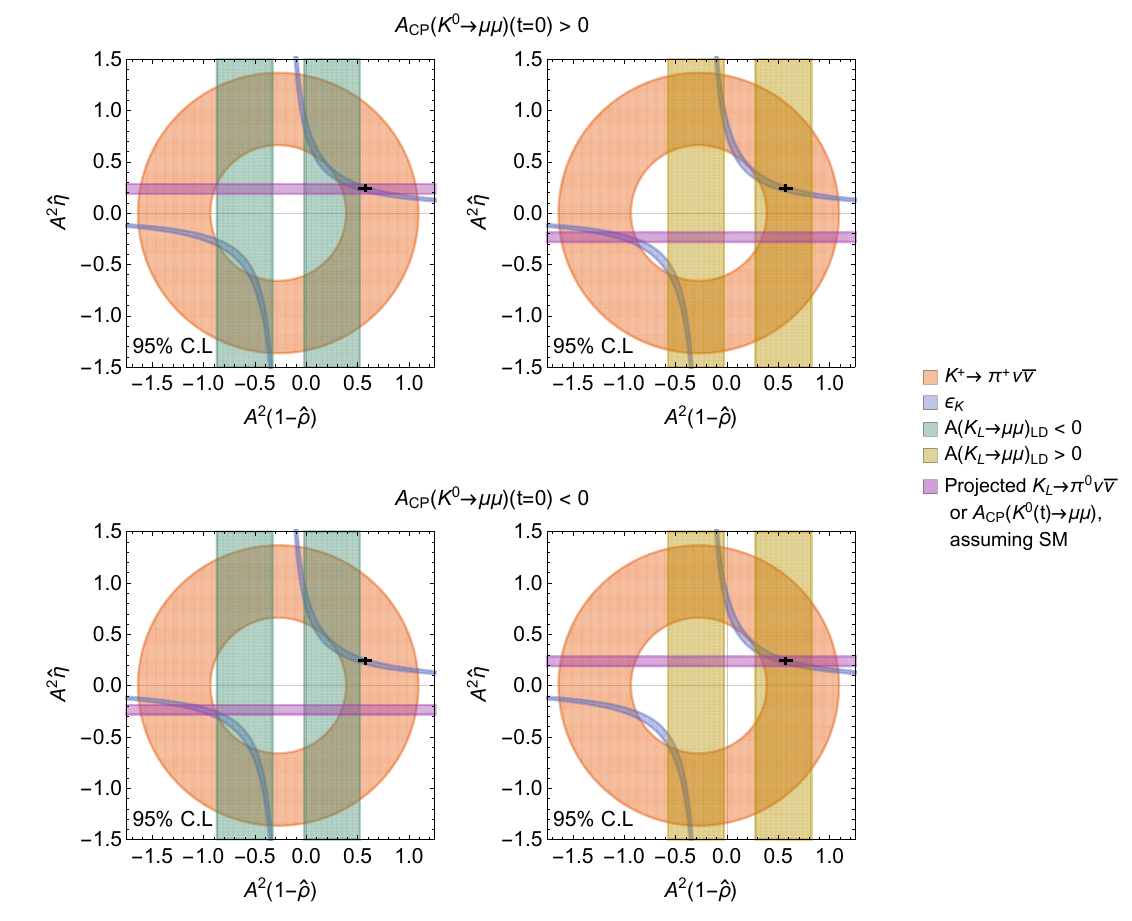}
    \caption{Supplemental Figure: Current ${\cal B}(K^+\to\pi^+\nu\bar\nu)$, $|\varepsilon_K|$ and ${\cal B}(K_L\to\mu^+\mu^-)$ constraints (as in Fig.~\ref{fig:2sigmaKLmumu}), overlaid with future projections for the measurement of $(A^2\hat\eta)$, assuming agreement with the SM. The top(bottom) panels corresponds to measuring a positive(negative) $A_{\rm CP}(K^0\to\mu^+\mu^-)$ at $t\approx0$.}
    \label{fig:PlAll}
\end{figure*}

We note that the measurement of $A_{\rm CP}(K^0\to\mu^+\mu^-)$ holds an additional piece of information, in its sign. The sign of $A_{\rm CP}(K^0\to\mu^+\mu^-)(t\approx 0)$ will determine the relative sign between the real part of the long distance amplitude, ${\cal R}eA_{\rm LD}$, and the short distance amplitude, proportional to $(A^2\lambda^4\hat\eta)$~\cite{Dery:2022yqc,Dery:ToAppear}.
This implies that only one of the two bands in Fig.~\ref{fig:FutureSM} ($\hat\eta>0$, $\hat\eta<0$) would go with each of the two interpretations of ${\cal B}(K_L\to\mu^+\mu^-)$ in Fig.~\ref{fig:2sigmaKLmumu} (see Figure~\ref{fig:PlAll} for the overlay of these two measurements, assuming a future SM-consistent measurement of $A_{\rm CP}(K^0\to\mu^+\mu^-)$, including its sign).
Then, if 
\begin{equation}
    {\rm sign}\left[{\cal R}eA_{\rm LD}\right] = {\rm sign[\hat\eta]}
\end{equation}
is determined and $|A_{\rm CP}(K^0\to\mu^+\mu^-)|$ is consistent with the SM, then even with current uncertainty on ${\cal B}(K_L\to\mu^+\mu^-)_{\rm SM}$, the "left island", as well as the "top" and "bottom" islands would be disfavored and we would be left with precision measurements to probe inside the "SM island".

\vskip0.5cm
{\bf Discussion}\quad
Traditional depictions of kaon information in the $(\bar\rho,\bar\eta)$ plane suffer not only from artificial error inflation, but also implicitly depend on B physics input, in the form of $|V_{cb}|$ dependence.
Therefore, the choice of a more natural plane for the presentation of kaon constraints elucidates the strength of the current and expected independent kaon physics picture, testing the CKM paradigm across sectors. 

Of the four parameters of the CKM matrix, three parameters are currently determined by B physics (excluding $|V_{us}|$). We demonstrate, that a different set of three parameters can in principle be determined by kaon physics (excluding $|V_{cb}|$). 
Putting $|V_{us}|$ and $|V_{cb}|$ aside (one determined and one undetermined within each sector), the remaining two parameters can be fixed by any of the normalized unitarity triangles, and can be depicted on a complex plane.
This framing makes it clear that, while it is natural to plot B physics constraints in the $(\bar\rho,\bar\eta)$ plane, characterizing the $(bd)$ unitarity triangle, 
the natural plane for kaon constraints is given by the $(ds)$ unitarity triangle, characterized by 
\begin{equation}
    -R_{ct}e^{-i\theta_{ct}}\, .
\end{equation}

The current status of kaon CKM determinations using this plane, as depicted in Figs.~\ref{fig:2sigmaCurrent} and~\ref{fig:2sigmaKLmumu}, exhibits the strengths of the community --- conquering the measurement of some of the rarest processes in particle physics, and providing theory predictions that use various advanced methods to control non-perturbative effects. The basic current picture is characterized by four disconnected allowed regions in parameter space, approximately symmetric around $\hat\eta=0$ and $A^2(1-\hat\rho)\approx -0.3$.

Upcoming theory and experimental progress will inevitably rule out some of these allowed "islands", enhancing our knowledge of the validity of the CKM paradigm in a fundamental way.
The analysis presented here suggests, that with upcoming advancements, either the "SM island" is the only region that will persist, or significant tension with the SM will be discovered.
In the scenario where only the "SM island" remains, any improvement in the experimental error on ${\cal B}(K^+\to\pi^+\nu\bar\nu)$ or the theory error on ${\cal B}(K_L\to\mu^+\mu^-)^{\rm SM}$ would directly impact the remaining allowed region, and we enthusiastically support any such efforts.

\vskip0.5cm
\textbf{\textit{Acknowledgments ---}}
We gratefully thank Stefan Schacht for careful reading of the manuscript and for meaningful discussions. We also thank Andrzej Buras, Yuval Grossman and Yossi Nir for their comments on the manuscript. 
A.D acknowledges support by the Weizmann Institute of Science Women's postdoctoral career development award.

%
%

\section{Appendix}
We summarize below the expressions used in our analysis, written in terms of $(A^2\lambda^4(1-\hat\rho),\,\, A^2 \lambda^4\hat\eta)$, as defined in Eq.~\eqref{eq:rhohatetahatDef}.
\begin{enumerate}[label=(\alph*)]
    \item {\bf $K^+\to\pi^+\nu\bar\nu$} \\ \vskip0.1cm
From Ref.~\cite{Buras:2015qea}, we have
\begin{eqnarray}\label{eq:KplusB}
    &\,&{\cal B}(K^+\to\pi^+\nu\bar\nu) = \widetilde\kappa_+ (1+\Delta_{\rm EM})X^2(x_t) \\ \nonumber
    &\,&\quad \times\left[{\cal I}m\left(\frac{V_{ts}^*V_{td}}{V_{cs}^*V_{cd}}\right)^2 + \left({\cal R}e\left(\frac{V_{ts}^*V_{td}}{V_{cs}^*V_{cd}}\right)+\frac{ \widetilde P_c}{X(x_t)} \right)^2\right] \\ \nonumber
    &\,&\qquad\qquad\quad\qquad = \widetilde\kappa_+(1+\Delta_{\rm EM})X^2(x_t)\\ \nonumber
    &\,&\quad \times\left[(A^2\lambda^4\hat\eta)^2 +\left(A^2\lambda^4(1-\hat\rho)+ \left(\frac{\widetilde P_c}{X(x_t)}\right)\right)^2\right]
\end{eqnarray}
where we use $\widetilde \kappa_+ = \kappa_+/\lambda^8$, $\widetilde P_c=P_c\cdot \lambda^4$, with~\cite{Buras:2015qea,Buchalla:1998ba}
\begin{align}
    \kappa_+ &= r_{K^+} \frac{3\alpha^2{\cal B}(K^+\to\pi^0 e^+\nu)}{2\pi^2 \sin^4\theta_W} \lambda^8 \\ \nonumber
    &= 5.17*10^{-11}\left(\frac{\lambda}{0.225}\right)^8\, , \\ \nonumber
    P_c &= \frac{1}{\lambda^4}\left[\frac{2}{3}X_{\rm NNL}^e +\frac{1}{3}X_{\rm NNL}^\tau\right] \\ \nonumber
    &= \left(0.404 \,\pm \, 0.024\right)\left(\frac{\lambda}{0.225}\right)^{-4}\, ,
\end{align}
and $X(x_t) = 1.481 \pm 0.009$.

\item {\bf $|\varepsilon_K|$}\\ \vskip0.1cm
From Ref.~\cite{Brod:2019rzc}, we have
\begin{eqnarray}
    |\varepsilon_K| &=& \kappa_\varepsilon C_\varepsilon \hat B_K\, \lambda^{2}\eta_{tt}{\cal S}_{tt}(x_c,x_t)\\ \nonumber
    &\times&A^2\lambda^4\hat\eta \left(A^2\lambda^4(1-\hat\rho) - \frac{\eta_{ut}{\cal S}_{ut}(x_c,x_t)}{\eta_{tt}{\cal S}_{tt}(x_c,x_t)}\right)\, ,
\end{eqnarray}
with $\kappa_\varepsilon=0.94\pm0.02$, $\hat B_K =0.7635\pm 0.0097$,
\begin{equation}
    C_\varepsilon = \frac{G_F^2 f_K^2 m_K m_W^2}{6\sqrt{2} \pi^2\Delta m_K}\, ,
\end{equation}
and where
\begin{eqnarray}
    \eta_{tt} = 0.55 (1\pm 4.2\%),\, \qquad \eta_{ut}=0.402(1\pm1.3\%)\, ,
\end{eqnarray}
and ${\cal S}_{tt}(x_c,x_t),\, {\cal S}_{ut}(x_c,x_t)$ are Inami-Lim functions.

\item {\bf $K_L\to\mu^+\mu^-$}\\ \vskip0.1cm
From Refs.~\cite{Isidori:2003ts,Hoferichter:2023wiy}, we have
\begin{eqnarray}
    {\cal B}(K_L\to\mu^+\mu^-) &=& {\cal B}(K_L\to\gamma\gamma)\cdot 2\beta_\mu \left(\frac{\alpha\, m_\mu}{\pi\, m_K}\right)^2  \\ \nonumber
    &\times&\left[{\cal I}mA_{\rm LD}^2 + \left({\cal R}eA_{\rm LD} + \chi_{\rm short}\right)^2\right]
\end{eqnarray}
with 
\begin{eqnarray}
    \chi_{\rm short} &=& \widetilde\kappa\cdot Y(x_t)\lambda\left({\cal R}e\left(\frac{V_{ts}^*V_{td}}{V_{cs}^*V_{cd}}\right)+ \frac{Y_{\rm NL}}{Y(x_t)}\right)\\ \nonumber
    &=& \widetilde\kappa\cdot Y(x_t)\lambda\left(A^2\lambda^4(1-\hat\rho)+ \frac{Y_{\rm NL}}{Y(x_t)}\right)\, , 
\end{eqnarray}
and we have $\tilde\kappa = 4988$, $Y(x_t)=0.95$, $Y_{\rm NL}=2.8\cdot 10^{-4}$, ${\cal I}mA_{\rm LD} = -5.2$, and~\cite{Hoferichter:2023wiy}
\begin{equation}
    {\cal R}eA_{\rm LD} = \begin{cases}
        0.16 \pm 0.38  \\ 
        -0.66 \pm 0.38 
    \end{cases}
\end{equation}

\item {\bf $K_S\to(\mu^+\mu^-)_{\ell=0}$}\\ \vskip0.1cm
From Ref.~\cite{Brod:2022khx}, we have
\begin{eqnarray}
    &\,&{\cal B}(K_S\to\mu^+\mu^-)_{\ell=0}     \\ \nonumber
    &\,& = \frac{\beta_\mu \tau_S}{16\pi m_K}\left|\frac{2\,G_F^2 m_W^2}{\pi^2}f_K m_K m_\mu Y_t\right|^2 \left|{\cal I}m\left(\frac{V_{ts}^*V_{td}}{V_{cs}^*V_{cd}}\right)\right|^2\\ \nonumber
    &\,& = \frac{\beta_\mu \tau_S}{16\pi m_K}\left|\frac{2\,G_F^2 m_W^2}{\pi^2}f_K m_K m_\mu Y_t\right|^2 (\lambda\cdot A^2\lambda^4\hat\eta)^2\, . 
\end{eqnarray}

\end{enumerate}

\bibliographystyle{apsrev4-1}
\bibliography{refs}

\end{document}